\documentclass[pra,nofootinbib]{revtex4}

\usepackage{graphicx}
\usepackage{amssymb}
\usepackage{amsmath}
\usepackage{amsfonts}
\usepackage{epstopdf}
\usepackage{epsfig}
\usepackage{wrapfig}

\begin{document}

\title{Quantum features of macroscopic fields. Entropy and dynamics}
\author{Robert Alicki}
\affiliation{International Centre for Theory
of Quantum Technologies (ICTQT), University of Gda\'nsk, 80-308, Gda\'nsk, Poland }

\begin{abstract}
Macroscopic fields like electromagnetic, MHD, acoustic or gravitational waves are usually described by classical wave equations with possible additional damping terms and coherent sources. The aim of this paper is to develop a complete macroscopic formalism including random/thermal sources, dissipation and random scattering of waves by  environment. The proposed \emph{reduced state of field} (RSF) combines averaged field with the two-point correlation function called single-particle density matrix. The evolution equations for RSF is obtained by reduction of the generalized quasi-free dynamical semigroups describing irreversible evolution of bosonic quantum field  and the definition  of RSF's entropy  follows from the von Neumann entropy of quantum field states. The presented formalism can be applied, for example, to superradiance phenomena and allows to unify the Mueller and Jones calculi in polarization optics.

\end{abstract}

\maketitle

\section{Introduction}
It is generally believed that low frequency waves appearing in the macroscopic world like 
various types of mechanical waves including acoustic ones, MHD, radio-frequency electromagnetic  or gravitational waves can be successfully described using classical wave equations with external sources \cite{Thorne}. This is certainly true for coherent deterministic sources while the case of thermal and generally random sources is much more problematic. It is not obvious how to incorporate classical waves as a part of thermodynamical system where exchange of heat, entropy production and generation of work should be taken into account. A simple addition of damping terms to wave equations is not sufficient. In particular the question of defining entropy for macroscopic fields and appropriate formulation of the Second Law is still an open problem. On the other hand it is clear that the proper description of physical systems is given by the quantum theory. However, the full formalism of the quantum field theory is too complicated and not convenient for practical applications. In most cases the relevant observables (energy, mass, momentum, angular momentum, polarization) are given by quadratic forms of fields and the linear approximation for dynamical equations is sufficient or can be easily amended by self-consistent non-linear corrections. This is similar to the case of classical gas where the  description in terms of particle density in the single-particle phase-space and the dynamics given by (linear) Boltzmann equation is more useful than the complete N-body formalism. In the case of macroscopic fields an approach based on single-particle density matrix (SPDM) and averaged field combined with the proper evolution equations is proposed, generalizing previous framework of quasi-free quantum dynamical maps and semigroups \cite{Alicki:78} - \cite{Alicki:07}. It includes processes of linear damping/pumping and random scattering of waves by the environment. In the case of purely reversible processes these master equations reduce to standard wave equations. This approach called reduced state of field  (RSF) formalism allows to use as thermodynamics entropy the formula obtained from the von Neumann entropy computed for the Gaussian state of the quantum field consistent with a given RSF.
\par
In order to illustrate in the simplest way the quantum features of classical fields we begin with the discussion of light polarization in terms of Stokes parameters and its quantum-mechanical interpretation. Then we present a complete description of a quantum field in terms of modes and its    \emph{First Quantization}  interpretation, where classical field is treated as a (not normalized)  quantum wave function of the corresponding (quasi) particle. The further step called \emph{Second Quantization} allows to describe irreversible processes in terms of Markovian Master Equations for density matrices acting of the corresponding bosonic Fock space. In the final step we develop the RSF formalism and apply it to the case of thermal sources and polarization optics.

\section{Jones and Mueller formalism for  light polarization} 

In optics, polarized light can be described using the Jones calculus, while partially polarized one is treated using Mueller calculus \cite{Stokes}.

Consider first a monochromatic plane wave of light propagating along axis - 3 in a Cartesian frame with the basis $\hat {\epsilon }_{k}$, $k = 1,2,3$. The (pure) state of polarization is specified by the complex amplitudes of the electric field, ${\displaystyle E_{1}}$ and ${\displaystyle E_{2}}$, in a basis ${\displaystyle ({\hat {\epsilon }}_{1},{\hat {\epsilon }}_{2})}$  The pair ${\displaystyle (E_{1},E_{2})} $ is called a \emph{Jones vector} and contains both, amplitudes and phases of two orthogonal components of the wave electric field. The relevant parameter is the normalization of the Jones vector 
\begin{equation}
s_0 =  |E_1|^2 + |E_2|^2
\label{S0}
\end{equation}
which is proportional to the energy density of the wave, hence also to the intensity of light beam $I$, and using quantum picture of light to the averaged photon number $N$. All those interpretations of $s_0$ can be useful in applications. With a given Jones vector we can associate a $2\times 2$ complex-valued matrix with matrix elements   $ E_{k} \bar{E}_{\ell}$. The  main idea of Stokes was to average those matrix elements with respect to fluctuations due either to slow fluctuations in time or the contribution to the light beam from different uncorrelated sources. Such a \emph{Stokes matrix} $\hat{S}$ can be decomposed with respect to Pauli matrices 
$\{\hat{\sigma}_k , k =1,2,3\}$ as
\begin{equation}
\hat{S} \equiv [ \langle E_{k} \bar{E}_{\ell}\rangle ] = \frac{1}{2}\sum_{\mu = 0}^3 s_{\mu} \hat{\sigma}_{\mu}
\label{Stokes}
\end{equation}
with $\hat{\sigma}_0 \equiv \mathbf{1}$ is the unit matrix. 
The real parameters $s_{\mu}$  are called \emph{Stokes parameters} and form a 4-dimensional \emph{Stokes vector} $\mathbf{s} \equiv (s_0 , \vec{s})$.  As the Stokes matrix $\hat{S}$ is positively defined the Stokes vector satisfies inequality
\begin{equation}
s_0 ^2 \geq s_1^2 + s_2^2 + s_3^2 .
\label{Stokes_1}
\end{equation}
Stokes anticipated that those parameters provide a \emph {complete description of  polarization state} of the monochromatic light beam. This assumption lies behind the Mueller calculus which describes the action of any \emph{linear} optical device by
$4\times 4$ Mueller matrix $\mathcal{M} = [M_{\mu\nu}]$ acting on the input Stokes vector $\mathbf{S}$ and yielding the output one 
$\mathbf{S}'$
\begin{equation}
 s_{\mu}' = \sum_{\nu = 0}^s  {M}_{\mu\nu}\, s_{\nu} .
\label{Mueller}
\end{equation}
Although, the Stokes matrix/vector is constructed from classical correlations between components of the classical electric fields the above completeness assumption is not consistent with the classical probabilistic scheme. Namely, treating polarization as a classical dynamical variable each fully polarized state of light satisfying the equality in \eqref{Stokes_1} with a fixed intensity $s_0$ corresponds to a pure state of the system. The set of all such pure states form the so-called \emph{Poincare sphere}. Therefore a classical mixed state of polarization corresponding to a partially or completely polarized monochromatic light beam  with a given intensity should be described by a probability measure on the Poincare sphere. Hence, the set of all mixed states is an infinite-dimensional simplex of all probability measures on the Poincare sphere generated by extreme points - the Dirac measures concentrated on all points of the sphere. On the other hand
in Stokes formalism any mixed state of polarization is given by the 3-dimensional vector with the length smaller or equal to $s_0$. This is completely equivalent to the description of quantum mixed states for the 2-level systems  with Poincare sphere replaced by \emph{Bloch sphere}. Therefore, one can say that Stokes was the first who discovered quantum nature of light, but similarly to Columbus was not aware of the meaning of his discovery \cite{Alicki:09}.
\par
The equivalence mentioned above sheds a new light on the Jones and Mueller calculi, which are useful tools in polarization optics.
Namely, using the well-known results from the quantum theory of open systems, \cite{Alicki:07}, \cite{Breuer}, \cite{Huelga} we can assume that any Mueller matrix corresponds to a completely positive map $\Phi$ such that the equation \eqref{Mueller} is equivalent to
\begin{equation}
\hat{S}' = \Phi (\hat{S}) =\sum_{\alpha} \hat{V}_{\alpha} \hat{S} \hat{V}_{\alpha}^{\dagger} 
\label{MuellerCP}
\end{equation}
where, the $2\times 2$ complex matrices $\hat{V}_{\alpha}$ are not uniquely defined, but one can always find the representation of $\Phi$ with at most 4 such matrices. The special case is a map $\Phi$ given by a single matrix $\hat{V}$ what means that each completely polarized state is transformed into another completely polarized one, albeit with different intensity. Therefore, one can restrict the description to map  $\hat{V}$ acting on Jones vectors, what is the essence of \emph{Jones calculus}. 
\par
The general Mueller map is completely positive but not trace preserving as $\mathrm{Tr}(\hat{S}) = s_0 $ and the intensity can change under the action of linear optics device. In principle, this theory describes not only absorption of light by the passive devices but also its amplification by active medium.
\par
One can ask what are the additional restrictions on the Mueller map $\Phi$ imposed by the Second Law of Thermodynamics. In the case of passive elements one can argue that the completely depolarized light can be transformed into polarized one only at the expense of intensity, i.e. the following condition must hold
\begin{equation}
 \Phi (\mathbf{1}) \leq \mathbf{1}, \quad \mathrm{or\ equivalently} \quad  \sum_{\alpha} \hat{V}_{\alpha} \hat{V}_{\alpha}^{\dagger} \leq \mathbf{1}.
\label{Mueller1}
\end{equation}
The natural question, related to thermodynamic properties of polarized light is the definition of entropy for a monochromatic light beam with a given polarization state described by $\hat{S}$. This question will be discussed in the Section 4 after generalization of Stokes formalism to other degrees of freedom of the macroscopic field.

\section{First and second quantization of classical field}

In this paper we restrict ourselves to the classical field occupying a finite volume and hence described by the set of complex modes $f_k(x)$, where $x$ is the position vector and $k$ denotes a discrete index. The modes  evolve in time according to the formula
\begin{equation}
f_k(x;t) = e^{-i\omega_k t} f_k(x)
\label{modes}
\end{equation}
and the arbitrary solution of the corresponding wave equation  can be represented by linear combination of modes.

In the picture of \emph{first quantization} modes $\{f_k\}$ correspond to (generally not normalised) energy eigenstates of the single-particle Hamiltonian $\hat{h}$ describing a single (quasi) particle associated with the field (e.g. photon, graviton, phonon, magnon, etc.).  The quantum single-particle Hilbert space $\mathcal{H}$ is spanned by those modes, with a proper normalization such that: \\
1) $\{f_k\}$ - form an orthonormal basis in $\mathcal{H}$,\\
2) classical energy of the field mode $f_k$ equals to $\hbar\omega_k$.

From now on we identify the classical field configuration represented by the linear superposition of modes with the corresponding vector in the Hilbert space of first quantization. Therefore, the  only mathematical difference between classical field and first quantization interpretation is the chosen normalization. In the first case we normalize field to the given energy or intensity, in the second to one treating classical field as a wave function of a single particle.

The \emph{second quantization} formalism describes the quantum field in terms of bosonic  Fock space $\mathcal{H}_F$ with a set of annihilation and creation operators $\{\hat{a}_{k} ,\hat{a}_k^{\dagger} \}$ corresponding to the modes $\{f_k\}$. They satisfy the canonical commutation relations
\begin{equation}
[\hat{a}_{k} ,\hat{a}_{k'}] = [\hat{a}^{\dagger}_{k} ,\hat{a}_{k'}^{\dagger}]= 0, \quad [\hat{a}_{k} ,\hat{a}_{k'}^{\dagger}] = \delta_{kk'} .
\label{CCR}
\end{equation} 
The Fock space is spanned by the vectors obtained by application of all monomials in creation operators acting on the \emph{vacuum state}
\begin{equation}
\bigl(\hat{a}_{k_1}\bigr)^{n_1} \bigl(\hat{a}^{\dagger}_{k_2}\bigr)^{n_2}\dots \bigl(\hat{a}^{\dagger}_{k_m}\bigr)^{n_m} |0\rangle .
\label{Fock}
\end{equation} 
For any vector $|\alpha\rangle = \sum_k \alpha_k |k\rangle$ in the single-particle Hilbert space there exists a normalized vector (pure state) $|\alpha_F\rangle$ in the Fock space, called \emph{coherent state}, which is a joint eigenvector of all annihilation operators
\begin{equation}
\hat{a}_k|\alpha_F\rangle = \alpha_k|\alpha_F\rangle.
\label{cohdef}
\end{equation} 
The coherent state  can be explicitly written as 
\begin{equation}
|\alpha_F\rangle = \sum_{n=0}^{\infty} \frac{1}{\sqrt{n!}}\bigl(\hat{a}^{\dagger}[\alpha]\bigr)^n|0\rangle \quad \mathrm{where}\quad
 \hat{a}^{\dagger}[\alpha]\equiv\sum_k \alpha_k \hat{a}^{\dagger}_k ,
\label{cohdef1}
\end{equation} 
or, introducing the Weyl unitary operators $\hat{W}[\alpha] $ on the Fock space, as
\begin{equation}
|\alpha_F\rangle = \hat{W}[\alpha] |0\rangle, \quad \mathrm{where}\quad \hat{W}[\alpha] = e^{(\hat{a}[\alpha]-\hat{a}^{\dagger}[\alpha])}.
\label{cohdef2}
\end{equation} 
The coherent state $|\alpha_F\rangle$ can be treated as the quantum analog of the classical field $|\alpha\rangle$.

In the following we restrict ourselves to two classes of operators acting on the Fock space obtained by two different \emph{lifting procedures}
applied to operators acting on the single-particle Hilbert space.

The single particle observable $\hat{b}$ expressed in the basis $\{|k\rangle \equiv |f_k\rangle\}$ as 
\begin{equation}
\hat{b} = \sum_{k,k'} b_{kk'}|k\rangle \langle k'|  
\label{observable}
\end{equation} 
produces an \emph{additive observable} on the Fock space $\cal{H}_F$ of the form
\begin{equation}
\hat{B} = \sum_{k,k'} b_{kk'} \hat{a}^{\dagger}_{k}\hat{a}_{k'} .
\label{observableF}
\end{equation} 
In particular for the Hamiltonian we have 
\begin{equation}
\hat{h} = \hbar \sum_{k} \omega_k |k\rangle \langle k|, \quad  \hat{H} = \hbar \sum_{k} \omega_{k} \hat{a}^{\dagger}_{k}\hat{a}_{k} \, ,
\label{hamF}
\end{equation} 
and the number operator is denoted by $\hat{N} = \sum_{k} \hat{a}^{\dagger}_{k}\hat{a}_{k} $.

Any unitary operator $\hat{u}$ acting on the single-particle Hilbert space extends to the Fock space \emph{multiplicative unitary} $\hat{\mathbf{U}}$ which can be defined in two equivalent ways
\begin{equation}
\hat{u} = e^{i\hat{b}} , \quad \hat{\mathbf{U}} = e^{i\hat{B}}, 
\end{equation} 
or
\begin{equation}
\hat{\mathbf{U}}\hat{C} \hat{\mathbf{U}}^{\dagger} = \hat{D} ,\quad \mathrm{where} \quad \hat{d} = \hat{u}\hat{c}\hat{u}^{\dagger}.
\label{unitary1}
\end{equation} 
The action of $\hat{\mathbf{U}}$ can be also defined in terms of coherent states or Weyl unitaries 
\begin{equation}
\hat{\mathbf{U}}|\alpha_F\rangle = |(u\alpha)_F\rangle, \quad  \hat{\mathbf{U}}\hat{W}[\alpha]\hat{\mathbf{U}}^{\dagger} = \hat{W}[u\alpha]
\label{unitarycoh}
\end{equation} 
where  $|u\alpha\rangle \equiv \hat{u}|\alpha\rangle $.

For the non-interacting field with dynamics governed by linear field equations with possible external classical and coherent sources the fundamental measurable quantities, like energy, momentum and angular momentum are additive observables. Therefore, instead of the full density matrix $\hat{\rho}_F$ acting of the Fock space $\cal{H}_F$ we can use the \emph{single-particle density matrix} (SPDM) $\hat{\rho}$ acting on the single-particle Hilbert space $\cal{H}$. The reduction map $\hat{\rho}_F \Rightarrow \hat\rho$ satisfies the following conditions
\begin{equation}
\mathrm{Tr} (\hat{\rho}_F \hat{B}) = \mathrm{tr} (\hat{\rho} \hat{b}) .
\label{observable1}
\end{equation} 
and 
\begin{equation}
\hat{\mathbf{U}}\hat{\rho}_F \hat{\mathbf{U}}^{\dagger} \Rightarrow   \hat{u}\hat{\rho}\hat{u}^{\dagger}.
\label{unitary2}
\end{equation} 
Here the trace $\mathrm{Tr}$ always refers to the Fock space  $\mathcal{H}_F$ and $\mathrm{tr}$ to the $\mathcal{H}$.

The explicit form of SPDM is given by 
\begin{equation}
\hat{\rho} = \sum_{k,k'} \mathrm{Tr} (\hat{\rho}_F \hat{a}^{\dagger}_{k'}\hat{a}_{k})|k\rangle \langle k'| ,
\label{exp}
\end{equation} 
which can be treated as the generalization of the idea of Stokes matrix to other degrees of freedom, beyond polarization. Notice that SPDM is normalized to the averaged number of particles, 
\begin{equation}
\mathrm{tr}\hat{\rho} = N = \mathrm{Tr}(\hat{\rho}_F \hat{N}).
\label{SPDM_norm}
\end{equation} 
The additional information about the phases of the field is contained in the \emph{averaged field} $|\alpha\rangle$ which is a vector in the single-particle Hilbert space $\cal{H}$ defined as
\begin{equation}
|\alpha\rangle = \sum_k\mathrm{Tr} (\hat{\rho}_F \hat{a}_k) |k\rangle  ,
\label{field}
\end{equation} 
generalizing the idea of Jones vector.
 
The definitions \eqref{exp} and \eqref{observableF} imply  that the \emph{correlation matrix} given by the formula
\begin{equation}
\hat{\rho}^{\alpha} = \hat{\rho} - |\alpha\rangle\langle\alpha| \geq 0
\label{correlation}
\end{equation} 
is a positive operator on the single-particle Hilbert space.

The \emph{reduced description} in terms of the pair $(\hat{\rho} , |\alpha\rangle)$ called \emph{reduced state of the field} (RSF) contains sufficient information about the most important properties of the macroscopic field interacting with environment. 
The RSF is called \emph{pure} if $\hat{\rho} = |\alpha\rangle\langle\alpha|$ or, equivalently $\hat{\rho}^{\alpha} = 0$. 
One can easily prove that the RSF is pure if and only if the original state of the quantum field is coherent.

\section{Quantum entropy of macroscopic field}

In phenomenological thermodynamics of equilibrium systems entropy is a function of a macroscopic state which is characterized by a small number of controlled  external parameters and temperature. Already in this case we have a certain freedom in selecting those external parameters related to our ability of controlling the system. The situation is more complicated for non-equilibrium systems where, typically, thermodynamic parameters including temperature become position-dependent and their choice is determined  by the relevant time-scales of local equilibration processes.

Similar problem appears when the notion of entropy is discussed within classical or quantum statistical mechanics. The proper choice of the definition depends on the selected level of complexity of our theoretical framework. This level is determined by the set of accessible observables of the system which can be measured and/or controlled. Again this choice is also related to relevant time-scales. Such          ``subjectivity" in the definition of entropy does not lead to any inconsistencies. Namely, the basic thermodynamical quantity with direct physical interpretation depending on entropy is the \emph{free energy} $F = U - T S$ ($U$-internal energy, $T$-temperature , $S$-entropy) which determines the amount of work extractable from the system. Obviously, both extractable work and entropy depend on the available means of control over the system.

In order to illustrate the problem of selection of complexity level we consider the classical gas of $N$ identical particles in a finite volume.The complete microscopical and statistical description of the state of such system is given by the $N$-particle probability distribution  $p_N(\vec{r}_1,\vec{p}_1 ,\dots ,\vec{r}_N,\vec{p}_N)$ which is symmetric with respect to permutations. 

The natural choice for the entropy of such state is the Gibbs expression
\begin{equation}
S^G (p_N) = -k_B \int\dots\int p_N(\vec{r}_1,\vec{p}_1 ,\dots ,\vec{r}_N,\vec{p}_N  ) \log p_N(\vec{r}_1,\vec{p}_1 ,\dots ,\vec{r}_N,\vec{p}_N  )\, d^3\vec{r}_1\, d^3\vec{p}_1 ,\dots ,d^3\vec{r}_N\, d^3\vec{p}_N
\label{Gibbsent}
\end{equation} 
However, the typical means of control over the gas are based on additive observables which does not involve correlations between individual particles. Therefore, for practical purposes, the statistical description of  gas in terms of marginal single-particle probability distribution $p(\vec{r},\vec{p})$ is sufficient. The standard definition of entropy in this case is the Boltzmann one, used in his description of gas dynamics 
\begin{equation}
S^B = -k_B N\int p(\vec{r},\vec{p}) \log p(\vec{r},\vec{p})\, d^3\vec{r}\, d^3\vec{p}
\label{Bent}
\end{equation} 
which coincide with  \eqref{Gibbsent} in the case of product probability distribution $p_N = \sqcap_{j=1}^N p(\vec{r}_j,\vec{p}_j)$.

Among the $N$-particle probability distributions with the same marginal $p(\vec{r},\vec{p})$ the product distribution maximizes the Gibbs entropy. Therefore, the Boltzmann choice \eqref{Bent} can be treated as the instance of the \emph{Maximum Entropy Principle} applied to the single-particle reduced description \cite{Ingarden}.

We follow the analogous reasoning for the case of macroscopic field described in terms of RSF $(\hat{\rho}, |\alpha\rangle)$. 
Consider first the \emph{quasi-free state} on the Fock space, generated by the additive observable $\hat{R}$ corresponding to the single-particle observable $\hat{b}$, which has form
\begin{equation}
\hat{\rho}'_F =  \frac{e^{-\hat{R}}}{\mathrm{Tr}{e^{-\hat{R}}}} .
\label{qfree}
\end{equation} 
One can easily compute the RSF corresponding to the state \eqref{qfree} obtaining
\begin{equation}
\hat{\rho}' =  \frac{1}{e^{\hat{r}} -1}, \quad |\alpha'\rangle = 0 ,
\label{qfree1}
\end{equation} 
and its von Neumann entropy
\begin{equation}
S_{vN} [\hat{\rho}'_F] = -k_B \mathrm{Tr}(\hat{\rho}'_F \log \hat{\rho}'_F ) = k_B \mathrm{tr}\bigl( (\hat{\rho}' +1) \log (\hat{\rho}' +1) - \hat{\rho}'  \log \hat{\rho}' \bigr) .
\label{qfree_ent}
\end{equation} 
To include also  macroscopic coherence one can apply the Weyl unitary transformation to produce the new state  
\begin{equation}
\hat{\rho}_F =  \hat{W}[\alpha]\hat{\rho}_F\hat{W}^{\dagger}[\alpha] .
\label{qfrees}
\end{equation} 
The RSF corresponding to the state \eqref{qfrees} is now $(\hat{\rho} , |\alpha\rangle)$ with
\begin{equation}
\hat{\rho} = \hat{\rho}' + |\alpha\rangle\langle\alpha|.
\label{qfrees1}
\end{equation} 
It is not difficult to check  that $\hat{\rho}_F $ of the form \eqref{qfrees} maximizes von Neumann entropy among all states on the Fock space with the given ($\hat{\rho}, |\alpha\rangle$). Therefore, we can define the entropy of RSF by the von Neumann entropy $S_{vN} [\hat{\rho}_F] = S_{vN} [\hat{\rho}'_F]$ what yields the expression depending on the correlation matrix $\hat{\rho}^{\alpha} =\hat{\rho} - |\alpha\rangle\langle\alpha| = \hat{\rho}'  $
\begin{equation}
S[\hat{\rho};|\alpha\rangle]   = k_B \mathrm{tr}\bigl( (\hat{\rho}^{\alpha} +1) \log (\hat{\rho}^{\alpha} +1) - \hat{\rho}^{\alpha}  
\log \hat{\rho}^{\alpha} \bigr) .
\label{SPDM_ent}
\end{equation} 
This entropy satisfies the natural conditions: $S[\hat{\rho};|\alpha\rangle] \geq 0$ and $S[\hat{\rho};|\alpha\rangle] = 0$ if and only if $\hat{\rho}=|\alpha\rangle\langle\alpha|$, i.e. RSF is pure.

The fact, that only coherent states of the quantum field produce zero entropy RSFs and all other pure states on the Fock space do not, illustrates the dependence of the notion of entropy on complexity of description which is determined by the assumed level of control.

\section{Generalized quasi-free dynamics}

The so-called quasi-free dynamical semigroups are completely positive trace preserving semigroups of dynamical maps acting on Fock space density matrices, which from the physical standpoint describe particle decay and production processes in the approximation of independent particles. In the following we introduce a more general class of such dynamical semigroups which include, additionally, coherent classical source and  individual and random scattering by the environment. It is assumed that a single scattering process is unitary, hence does not produce a persistent entanglement with environment. The  master equation satisfying the assumptions of above takes form ($\{\cdot , \cdot\}$ denotes anticommutator)
\begin{align}
\frac{d}{dt}	\hat{\rho}_F	&= -i\hbar \sum_{k} \omega_{k} [\hat{a}^{\dagger}_{k}\hat{a}_{k}, \hat{\rho}_F	] + \sum_{k}  [(\xi_k\hat{a}^{\dagger}_{k} - \bar{\xi}_k \hat{a}_{k}), \hat{\rho}_F	] \nonumber \\
&+\sum_{k,k'} \Gamma_{\downarrow}^{kk'} \bigl(\hat{a}_{k} \hat{\rho}_F\hat{a}^{\dagger}_{k'} -\frac{1}{2}\{\hat{a}^{\dagger}_{k'}\hat{a}_{k}, \hat{\rho}_F\}	\bigr) \nonumber\\ 
&+\sum_{k,k'} \Gamma_{\uparrow}^{kk'} \bigl(\hat{a}^{\dagger}_{k} \hat{\rho}_F\hat{a}_{k'} -\frac{1}{2}\{\hat{a}_{k'}\hat{a}^{\dagger}_{k}, \hat{\rho}_F\}	\bigr) \nonumber\\
&+ \int \mu(du) \bigl( \hat{\mathbf{U}}\hat{\rho}_F \hat{\mathbf{U}}^{\dagger} - \hat{\rho}_F\bigr) .
\label{GMME}
\end{align}
In the formula of above the complex amplitudes $\xi_k$ describe the coherent source of field, the positive matrices $[\Gamma_{\downarrow}^{kk'}]$ and $[\Gamma_{\uparrow}^{kk'}]$ contain particle annihilation and production rates for random sources. Those rates are expressed as the eigenvalues of  $[\Gamma_{\downarrow}^{kk'}]$ and $[\Gamma_{\uparrow}^{kk'}]$, respectively. The last term describes random scattering processes parametrized by the positive measure $\mu(du)$, or more generally, positive distribution defined on the  group of all unitaries acting on $\mathcal{H}$. In particular, when the Poisson process of random scattering tends to its diffusion limit one obtains the \emph{double commutator} terms $ -[\hat{Q}, [\hat{Q}, \hat{\rho}_F]]$, with an additive observable $\hat{Q}$, often used to describe pure decoherence.

The equation \eqref{GMME} possesses the standard  Gorini-Kossakowski-Lindblad-Sudarshan form \cite{GKS} - \cite{Huelga} and hence its solution is given by the completely positive trace-preserving dynamical semigroup.

By direct calculation one can obtain from \eqref{GMME} the closed evolution equation for the reduced description of the field in terms of
the RSF $(\hat{\rho} , |\alpha\rangle)$.  Introducing the single-particle positive operators representing damping and pumping
\begin{equation}
\hat{\gamma_{\downarrow}} =  \sum_{k,k'}\Gamma_{\downarrow}^{kk'}  |k\rangle \langle k'| , \quad
\hat{\gamma_{\uparrow}} =  \sum_{k,k'}\Gamma_{\uparrow}^{kk'}  |k\rangle \langle k'| ,
\label{dampump}
\end{equation} 
and the single particle vector describing coherent source
\begin{equation}
|\xi\rangle = \sum_k \xi_k |k\rangle  
\label{source}
\end{equation} 
one can write the equations of motion in the form of two coupled equations which will be called \emph{reduced kinetic equations} (RKEs) for RSF
\begin{align}
\frac{d}{dt}	\hat{\rho}	&= -\frac{i}{\hbar} [\hat{h}, \hat{\rho}] + (|\xi\rangle\langle \alpha|+ |\alpha\rangle\langle\xi|) \nonumber \\
&+\{(\hat{\gamma}_{\uparrow} -\hat{\gamma}_{\downarrow}), \hat{\rho}\} + \hat{\gamma}_{\uparrow} \nonumber\\ 
&+ \int \mu(du) \bigl( \hat{u}\hat{\rho} \hat{u}^{\dagger} - \hat{\rho}\bigr) .
\label{MESPDM}
\end{align}

\begin{align}
\frac{d}{dt}|\alpha\rangle		&= -\frac{i}{\hbar}\hat{h} |\alpha\rangle + \frac{1}{2} \bigl(\hat{\gamma}_{\uparrow} -\hat{\gamma}_{\downarrow}\bigr)|\alpha\rangle
 +|\xi\rangle \nonumber \\
&+ \int \mu(du) \bigl( \hat{u} - 1)\bigr)|\alpha\rangle .
\label{MESPDM1}
\end{align}
The RKEs \eqref{MESPDM} and \eqref{MESPDM1} are exact consequences of the master equation  \eqref{GMME} describing linear quantum field dynamics, both reversible and dissipative. One can include also nonlinear effects on the level of RSF by introducing the dependence on 
$(\hat{\rho} , |\alpha\rangle)$ of the operators $\hat{h}, \hat{\gamma}_{\uparrow}, \hat{\gamma}_{\downarrow}$ and the measure $\mu(du)$ in the spirit of self-consistent Hartree-Fock approximation for bosons. Another generalization is needed to include the case of varying external conditions. This can be done by introducing also time-dependence into the operators $\hat{h}, \hat{\gamma}_{\uparrow}, \hat{\gamma}_{\downarrow}$,  the measure $\mu(du)$ and the coherent external source $|\xi\rangle$. In the absence of coherent source the equations become decoupled.

Notice, that only in the absence of random source $(\hat{\gamma}_{\uparrow} = 0)$ and random scattering $(\mu(du) = 0)$ and under the assumption that the initial RSF  is pure, i.e. $\hat{\rho}(0) = |\alpha(0)\rangle\langle\alpha(0)|$, it remains pure and satisfies classical field equation with damping and coherent source
\begin{equation}
\frac{d}{dt}|\alpha\rangle		= -\Bigl(\frac{i}{\hbar}\hat{h} + \frac{1}{2} \hat{\gamma}_{\downarrow}\Bigr)|\alpha\rangle
 +|\xi\rangle .
\label{MESPDM2}
\end{equation}
Although this type of equations is quite frequently used its applicability is limited to classical coherent sources, zero-temperature environment and the absence of random scattering.

\section{Examples}
In order to illustrate the presented formalism of RKEs for RSF, \eqref{MESPDM},\eqref{MESPDM1}, we consider two special cases: macroscopic field in thermal environment and linear polarization optics.

\subsection{Thermal environment}

The  RKEs proposed above with possible non-linear and time-dependent generalizations provide phenomenological tools to deal with macroscopic field interacting with an environment. There exist examples where special classes of RHEs can be derived from the underlying Hamiltonian models of the field interacting  with the thermal bath and using appropriate approximations.  The most popular approximation scheme combines Born , Markovian and secular ones and leads to the operators $\hat{\gamma}_{\uparrow} ,\hat{\gamma}_{\downarrow}$ diagonal in energy representation (we neglect also coherent source $|\xi\rangle = 0$)
\begin{equation}
\hat{\gamma_{\downarrow}} =  \sum_{k}\gamma_{\downarrow}(k)  |k\rangle \langle k| , \quad
\hat{\gamma_{\uparrow}} =  \sum_{k}\gamma_{\uparrow}(k)  |k\rangle \langle k| ,
\label{dampump1}
\end{equation} 
with the additional condition implied by the thermal character of the bath
\begin{equation}
\gamma_{\uparrow}(k) = e^{-(\hbar\omega_k/k_B T)} \gamma_{\downarrow}(k)  .
\label{kms}
\end{equation} 
The rates $\gamma_{\uparrow}(k), \gamma_{\downarrow}(k) $ can be computed using Fermi Golden Rule \cite{RA-Fermi}. The random scattering term can also be derived using the alternative low density limit for the suitable Hamiltonian model of field-- perturber elastic scattering. In this case the measure $\mu(du)$ in eqs.\eqref{MESPDM}, \eqref{MESPDM} is concentrated on unitaries commuting with $\hat{h}$. Under the above assumptions one obtains the independent set of equations for the diagonal elements of SPDM,  $n_k = \langle k|\hat{\rho}_1|k\rangle$ (describing particle occupation numbers)  and averaged field amplitudes $\alpha_k$
\begin{equation}
\frac{d}{dt} n_k		= -\bigl({\gamma}_{\downarrow}(k) -{\gamma}_{\uparrow}(k)\bigr) n_k  + {\gamma}_{\uparrow}(k) ,
\label{diag}
\end{equation}
\begin{equation}
\frac{d}{dt}\alpha_k	= -\Bigl(i\omega'_k + \frac{1}{2} \bigl(\gamma_{\downarrow}(k) - \gamma_{\uparrow}(k)\bigr) + \gamma_{dec}(k)\Bigr)\alpha_k
\label{fieldamp}
\end{equation}
with the \emph{decoherence rate} 
\begin{equation}
\gamma_{dec}(k) = \Re{\int \mu(du) \bigl(1 - \langle k|\hat{u}|k\rangle\bigr)} \geq 0.
\label{fieldec}
\end{equation}
Here, we assume that   $\Im{\int \mu(du) \bigl(1 - \langle k|\hat{u}|k\rangle\bigr)} $ is absorbed into renormalized frequency $\omega'_k$.

The equations \eqref{fieldamp} and \eqref{diag} can be seen as the manifestation of \emph{wave-particle duality} in the description of macroscopic field. The quantum-field feature of the system is hidden in the form of energy damping rates $(\gamma_{\downarrow}(k) - \gamma_{\uparrow}(k)\bigr)$ where the minus sign by the second term is a consequence of the \emph{stimulated emission} related to bosonic character of field excitations (particles). 

The quantum phenomenon of stimulated emission becomes particularly interesting for moving heat baths interacting with the macroscopic field. The case of rotating heat baths has been discussed in details in \cite{rotatingbath} where quantum master equations of the type \eqref{GMME} (with diagonal matrices $[{\Gamma_{\uparrow(\downarrow)}}^{kk'}]$ and absent coherent sources and random scattering) were used. The only consequence of bath rotation is the modification of the relation \eqref{kms} into
\begin{equation}
\gamma_{\uparrow}(k) = e^{-[\hbar(\omega_k - m(k)\Omega)/k_B T]} \gamma_{\downarrow}(k)  .
\label{kms1}
\end{equation} 
where $m(k)$ is a magnetic quantum number of the mode $|k\rangle$ and $\Omega$ is the angular frequency of rotation. The modes for which
$ \omega_k < m(k)\Omega $ possess a negative energy damping rate $({\gamma}_{\downarrow}(k) -{\gamma}_{\uparrow}(k)\bigr)$, what means that the kinetic energy of rotating bath is pumped into these modes. Moreover, if the negative damping of the averaged mode amplitude $({\gamma}_{\downarrow}(k) -{\gamma}_{\uparrow}(k)\bigr)/2$ dominates over the decoherence rate $\gamma_{dec}(k)$ those amplitudes are amplified. This phenomenon is called \emph{superradiance} and can be studied for various physical implementations: from Hawking radiation of rotating black holes to ocean wave generation by wind \cite{rotatingbath}, \cite{Bekenstein}, \cite{BCP}.

\subsection{Polarization optics revisited}

The presented above method of reduced description of quantum field can be applied to the polarization degrees of freedom in the case of linear optics devices. Namely, one can consider a light beam consisting of  photons occupying the modes with a narrow band of frequencies around the central frequency $\omega^0$ and with a fixed spatial structure. Therefore, the reduced description in terms of RSF is given by the $2\times 2$ positively defined Stokes matrix of the  Section 2, but now obtained from averaging over the full quantum state $\hat{\rho}_F$ of the light beam  
\begin{equation}
\hat{S} \equiv [S_{k\ell}], \quad S_{k\ell} = \sum_q\mathrm{Tr}\bigl(\hat{\rho}_F \hat{a}^{\dagger}_{\ell q}\hat{a}_{k q}\bigr), \quad k,\ell = 1,2 .
\label{Stokes1}
\end{equation}
Here, $q$ denotes the other then polarization quantum numbers of light beam modes. The transmission of the beam  from the entrance to the exit of the linear optics device can be treated as the time evolution governed by the master equation of type \eqref{MESPDM} with the absent coherent and incoherent sources (compare to the discussion of fiber optics in \cite{Benatti:2006}). This evolution equation for the Stokes matrix reads
\begin{equation}
\frac{d}{dt}	\hat{S}	= -i[\hat{\omega}, \hat{S}]
 -\{\hat{\gamma}_{\downarrow}, \hat{S}_1\}  
+ \int \mu(du) \bigl( \hat{u}\hat{S} \hat{u}^{\dagger} - \hat{S}\bigr) .
\label{MESPDMopt}
\end{equation}
where :

i) $\hat{\omega} $ is a Hermitian  $2\times 2$ matrix describing  rotation of polarization vector,

ii) $\hat{\gamma}_{\downarrow}$ is a positive $2\times 2$ matrix describing absorption of photons by the medium,

iii) $\hat{u}$ are  $2\times 2$ unitaries describing depolarization of light by random scattering with the positive weight $\mu(du)$.

From the discussion in the previous Sections it follows that the reduced description of a quantum field involves also the averaged field as an observable object. For  polarization of a light beam this is a 2-dimensional complex  vector $|\alpha\rangle = [\alpha_1 , \alpha_2]$ which is equivalent to the standard Jones vector with the normalization determined by the following definition
\begin{equation}
|\alpha\rangle = \sum_{k=1}^2\bigl[\sum_q\mathrm{Tr}\bigl(\hat{\rho}_F \hat{a}_{k q}\bigr)\bigr]|k\rangle .
\label{Jones1}
\end{equation}
The equation of motion for the averaged Jones vector is decoupled from \eqref{MESPDMopt}, but contains the same parameters 
\begin{equation}
\frac{d}{dt}|\alpha\rangle	= -\Bigl(i\hat{\omega}' + \frac{1}{2} \hat{\gamma}_{\downarrow} + \hat{\gamma}_{dec}\Bigr)|\alpha\rangle
\label{poldamp}
\end{equation}
with  $\hat{\omega}'= \hat{\omega}+ \hat{\delta}$, where
\begin{equation}
\int \mu(du) \bigl(1 - \hat{u}\bigr) = i\hat{\delta} + \hat{\gamma}_{dec}, \quad  \hat{\delta}=\hat{\delta}^{\dagger}, 
\quad  \hat{\gamma}_{dec} \geq 0 .
\label{decmat}
\end{equation}
Integrating the equation of motion \eqref{MESPDMopt} between entry and exit times we obtain a global dynamical map characterizing the influence of linear optics device on the polarization state. Because the first two  terms on the RHS of \eqref{MESPDMopt} generate a pure contracting CP maps and the third term generates bistochastic CP maps the most general Mueller map satisfies
\begin{equation}
\hat{S}' = \Phi (\hat{S}),\quad   \Phi (\mathbf{1}) \leq \mathbf{1}, \quad \Phi^* (\mathbf{1}) \leq \mathbf{1} ,
\label{MuellerCP1}
\end{equation}
where $\Phi^* $ is a dual (Heisenberg picture) map. Such CP map can be called \emph{doubly contracting}.
In terms of the explicit representation
\begin{equation}
\Phi (\hat{S}) =\sum_{\alpha} \hat{V}_{\alpha} \hat{S} \hat{V}_{\alpha}^{\dagger} , \quad \Phi^* (\hat{M}) =\sum_{\alpha} \hat{V}_{\alpha} ^{\dagger}\hat{M} \hat{V}_{\alpha}
\label{Mueller2}
\end{equation}
the conditions \eqref{MuellerCP1} can be written as
\begin{equation}
\sum_{\alpha} \hat{V}_{\alpha} \hat{V}_{\alpha}^{\dagger} \leq \mathbf{1},\quad \sum_{\alpha} \hat{V}_{\alpha} ^{\dagger}\hat{V}_{\alpha} \leq \mathbf{1} .
\label{Mueller3}
\end{equation}
The Kraus decomposition in \eqref{Mueller2} is not unique, but one can always choose at most 4 matrices $\hat{V}_{\alpha}$.

Similarly the corresponding  $2\times 2$ matrix acting on Jones vectors is contracting
\begin{equation}
|\alpha'\rangle =  \hat{V}|\alpha\rangle ,\quad \hat{V} \hat{V}^{\dagger} \leq \mathbf{1} .
\label{Jonespol}
\end{equation}
In the most general case the only condition which connects $\Phi$ and $\hat{V}$ is that for each pair of Stokes matrix $\hat{S}$ and Jones vector $|\alpha\rangle$ 
\begin{equation}
 \hat{S} \geq |\alpha\rangle\langle\alpha|\quad \mathrm{implies}\quad \Phi (\hat{S}) \geq 
\hat{V}|\alpha\rangle\langle\alpha|\hat{V}^{\dagger}.
\label{MuJon}
\end{equation}
The condition implies that the difference of two CP maps $ \Phi -\hat{V} \cdot \hat{V}^{\dagger}$ is positive. 

Summarizing, in contrast to a general believe that Jones and Mueller calculi refer to physically different situations we argue that the complete description of the polarization state of light beam needs a pair $(\hat{S}, |\alpha\rangle)$ of Stokes matrix and averaged Jones vector satisfying $\hat{S} \geq |\alpha\rangle\langle\alpha|$. Equivalently, one can use Stokes parameters and explicit components of Jones vector in the given polarization basis $[s_0 , \vec{s} ; \alpha_1 , \alpha_2]$.
The action of any linear optics device is described by a pair $(\Phi , \hat{V})$ of  CP and doubly contracting Mueller map acting on $2\times 2$ matrices and the Jones contracting $2\times 2$ matrix such that the map $\Phi - \hat{V}\cdot\hat{V}^{\dagger}$ is positive. The equivalent representation of $(\Phi , \hat{V})$ is given by a pair $\{[M_{\mu\nu}], \mu,\nu = 0,1,2,3; [V_{k\ell}, k,l =1,2\}$ of Mueller and Jones matrices. We call such a set of pairs in both representations \emph{Mueller-Jones maps}.

The set of Mueller-Jones maps form a semigroup with the  composition of two elements
\begin{equation}
(\Phi , \hat{V}) \circ (\Phi' , \hat{V'}) \equiv (\Phi\Phi', \hat{V}\hat{V'}),
\label{composition}
\end{equation}
which physically means a new optical device composed of two aligned ones.

Finally, we can settle down the question of entropy for polarization of a light beam using the definition \eqref{SPDM_ent}, which now takes form
\begin{equation}
S[\hat{S};|\alpha\rangle]   = k_B \mathrm{tr}\bigl( (\hat{S}^{\alpha} +1) \log (\hat{S}^{\alpha} +1) - \hat{S}^{\alpha}
\log \hat{S}^{\alpha}_1 \bigr) .
\label{pol_ent}
\end{equation} 
with $\hat{S}^{\alpha} \equiv \hat{S} -|\alpha\rangle\langle\alpha|$. 

\section{Concluding remarks}

The presented, mathematically consistent, formalism of reduced description of quantum fields has a potentially wide range of applications. It is rather surprising that quantum features, in particular quantum statistics, have such an influence on the macroscopic behavior of wave-like excitations. Even for such macroscopic objects like ocean  waves generated by wind or MHD waves in stellar atmospheres the stimulated emission processes  characteristic for bosons may lead to various macroscopic phenomena like superradiance or creation of shock waves. The description in a form of two mathematical objects -- averaged field and population numbers (diagonal elements of the single-particle density matrix) -- can be seen as a macroscopic manifestation of particle-wave duality in quantum world. Namely, for coherent sources, zero temperature environment and absent random scattering the description in terms of wave equations with sources and pure damping is sufficient. When random/thermal effects prevail the averaged field tends rapidly to zero and kinetic equations for (quasi) particle population numbers govern the evolution of the relevant observables. Such wave-particle transition in the modeling of physical phenomena may explain the origin of singularities in purely classical theories like hydrodynamics or general relativity.

\textit{Acknowledgments} 
The author was supported by the Foundation for Polish Science (FNP) through its International Research Agendas (IRAP), with structural funds from the European Union (EU) for the ICTQT.


\bibliographystyle{mdpi}

\begin{thebibliography}{10}

\bibitem{Thorne}
{K.S. Thorne and R.D. Blandford},
\newblock {\em Modern Classical Physics}, 
\newblock{Princeton University Press}, Princeton and Oxford, 2017.
  2007.

\bibitem{Alicki:78}
{R. Alicki },
\newblock {The Theory of Open Systems in Application to Unstable Particles},
\newblock {\em Rep. Math. Phys.}, 14:27, 1978.

\bibitem{Evans}
{D.E. Evans},
\newblock {Completely positive quasifree maps on the CAR algebra},
\newblock {\em Commun. Math. Phys.}, 70:53, 1979.	

\bibitem{Verbeure}
{B. Demoen, P. Vanheuverzwijn and A. Verbeure},
\newblock{Completely positive quasi-free maps of the CCR-algebra},
\newblock{\em Rep. Math. Phys.}, 15:27, 1979

\bibitem{Alicki:07}
{R. Alicki and K. Lendi},
\newblock{\em Quantum Dynamical Semigroups and Applications}, II-nd edition,
\newblock LNP 717. {Springer-Verlag}, Berlin,
  2007.

\bibitem{Stokes}
{S.N. Savenkov},  
\newblock{Jones and Mueller matrices: Structure, symmetry relations and information content},
\newblock{\em Light Scattering Reviews} 4:71,  2009.


\bibitem{Alicki:09}
{R. Alicki },
\newblock{On von Neumann and Bell Theorems Applied to Quantumness Tests}, 
\newblock{\em Foundations of Physics}, 39:352, 2009.

\bibitem{Ingarden}
{R.S. Ingarden, A. Kossakowski and M. Ohya},
\newblock {\em Information Dynamics and Open Systems},
\newblock {Kluwer Academic Publishers}, Dordrecht, 1997.
	
\bibitem{GKS}
{V. Gorini, A. Kossakowski and E.C.G. Sudarshan},
\newblock {Completely positive dynamical semigroups of n-level systems}.
\newblock {\em J. Math. Phys.}, 17:821, 1976.

\bibitem{Lindblad}
{G. Lindblad},
\newblock {On the generators of quantum dynamical semigroups},
\newblock {\em Commun. Math. Phys.}, 40:147, 1976.

\bibitem{Breuer}
{H.-P. Breuer and F. Petruccione},
\newblock {\em The Theory of Open Quantum Systems},
\newblock {Oxford University Press}, Oxford, 2002.

\bibitem{Huelga}
{A. Rivas and S.F. Huelga},
\newblock {\em  Open Quantum Systems. An Introduction},
\newblock {SpringerBriefs in Physics}, Heidelberg, 2012.

\bibitem{RA-Fermi}
{R. Alicki},
\newblock{The Markov master equation and the Fermi golden rule},
\newblock{\em	Int. J. Theor. Phys.},  16:351,  1977.

\bibitem{rotatingbath}
{R. Alicki and A. Jenkins},
\newblock{Interaction of a quantum field with a rotating heat bath},
\newblock{\em Ann. Phys. (NY)},  395: 69, 2018.
	

\bibitem{Bekenstein}
{J.D. Bekenstein and M. Schiffer},
\newblock{The many faces of superradiance},
\newblock{\em	Phys. Rev. D},  58: 064014 , 1998.
	[arXiv:gr-qc/9803033]
	

\bibitem{BCP}
{R. Brito, V. Cardoso and P. Pani},
\newblock{\em Superradiance: Energy extraction, black-hole bombs and implications for astrophysics and particle physics},
\newblock{	Lect. Notes Phys. 906}, Springer, Heidelberg, 2015.


\bibitem{Benatti:2006}
{F. Benatti and R. Floreanini},  
\newblock{Tests of Complete Positivity in Fiber Optics},
\newblock{\em Open Systems and Information Dynamics}, 13:229, 2006 


\end{thebibliography}

\end{document}